\documentclass[prc,showpacs,twocolumn]{revtex4}
\usepackage{graphicx}
\begin{document}

\title{Particle distribution and nuclear stopping in Au-Au collisions at
$\sqrt{s_{NN}}$=200 GeV}
\author{L.L.Zhu and C.B. Yang}
\affiliation{Institute of Particle Physics, Hua-Zhong Normal University, Wuhan 430079,
P.R. China}

\begin{abstract}
The transverse momentum distribution of produced charged particles is investigated
for gold-gold collisions at $\sqrt{s_{NN}}=200$ GeV. A simple parameterization
is suggested for the particle distribution based on the nuclear stopping
effect. The model can fit very well both the transverse momentum distributions
at different pseudo-rapidities and the pseudo-rapidity distributions at different
centralities. The ratio of rapidity distributions for peripheral and central
collisions is calculated and compared with the data.

\end{abstract}

\maketitle

\section{Introduction}
Ultra-relativistic heavy ion collision is the only way to study
nuclear matters at extremely high temperature and density in the
controlled experimental conditions. Such a study can provide
information about the internal structure of hadrons and the
interactions among partons from which those hadrons are formed.
An important quantity for characterizing the properties of the
produced particles is the transverse momentum distribution
in different rapidity regions. The transverse momentum distribution
at mid-rapidity region can tell us the matter density created
in heavy ion collisions and that at large rapidity tells us the baryon
stopping strength in the collisions. Experimentally, the charged
hadrons can be comparatively easily identified and investigated.
The multiplicity distribution of charged particle is often used
as a global measure of the dynamics of the ultra-relativistic heavy
ion collisions. The total multiplicity as a function of
rapidity or pseudo-rapidity can be used to measure the nuclear
stopping effect in the collisions. Because the stopping effect is directly
related to how effectively the kinetic energy of the colliding particles
can be converted into thermal one and how many secondaries can be produced
in the process, it has been a focus on both nucleon-nuclear and
nucleus-nucleus collisions for quite a long time and been studied in various
approaches before \cite{hwa84,csernai,huf,wong,kop,khar,van}. Some of the
investigations are done at the nucleon level, some others are based on the
string model. Since it is not feasible to perform a first-principle
parameter-free calculation of the nuclear stopping effect, experimental guidance
plays a crucially important role in such studies.

This paper discusses the transverse momentum distributions of emitted
particle at the highest energy of the Relativistic High Ion
Collider (RHIC) between gold-gold (Au-Au) at $\sqrt{s_{NN}}=200$
GeV for different centralities and rapidities. We will
start from a very simple parameterization for the particle distribution
stimulated from momentum degradation effect and fit the experimental data
of the transverse momentum distributions for charged particles for different
centralities and rapidities. Then the rapidity distributions can be obtained
from integration over transverse momentum for different centralities and
are compared to the experimental data.

The organization of the paper is as follows. In section II, we will
propose a simple parameterization for the invariant particle distributions.
Then in section III, we fit the experimental data to our model parameterization.
The last section is for a brief conclusion.

\section{Parameterization for the invariant particle distribution}
More than 30 years ago, the transverse momentum distribution
of particles produced at polar angle $\theta=\pi/2$ in
high energy proton-proton collisions was suggested \cite{car73} as
\begin{equation}
E\frac{d^3\sigma}{dp^3}=A\exp(-\xi^2/2b)\ ,
\end{equation}
from the analogy of Landau's hydrodynamical model,with $\xi=\ln (m_T+p_T)/m$
the transverse rapidity and $b$ a parameter to be determined by fitting the
experimental data. Here $m$ is the mass of the produced particle, ${m_T}=
\sqrt{m^2+p_T^2}$ is the transverse mass. Recently this parameterization
was also used in describing the data for nucleus-nucleus collisions at RHIC energy
\cite{nao05} in the mid-rapidity region. At small $p_T$ the above transverse momentum
distribution can be approximated by an exponential and the collective flow effect
can be extracted if the temperature of phase transition is known. So, The value
of $b$ can tell us some information about the collective flow. In recent years,
some authors have used the blast wave model to parameterize the transverse momentum
distribution, see \cite{toma} for example. That model is valid only at low $p_T$
region, since the blast wave is a collective dynamical phenomenon for the produced
hot medium and the contribution from hard partons to the particle production cannot
be contained in the blast wave model. For the distributions of particles with
nonzero longitudinal momentum $p_L$, no first-principle theory, such as
quantum chromodynamics, can be used to
derive a formula for the distribution. Therefore, a phenomenological ansatz is
needed. In this paper, as usually done in other studies, we simply assume a
factorized form for the distribution as
\begin{equation}
E\frac{d^3N}{dp^3}=A\exp(-\xi^2/2b)H(p_L)\ .
\end{equation}

In a former study \cite{hwa02} of $p-A$ collisions it is shown that the Feynman $x_F$
(or longitudinal momentum $p_L$) distribution can be approximated by an exponential form.
The exponential decrease of the longitudinal momentum distribution is a reflection
of the momentum degradation effect from nucleon-nucleon interactions while
incident nucleon is penetrating the target. For nucleus-nucleus collisions, one can
expect that the distribution may be not too far from that for $p-A$ collisions. Thus we
assume
\begin{equation}
H(p_L)\propto \exp(-cp_L^d)\ ,
\end{equation}
with $c, d$ two parameters characterizing the longitudinal momentum degradation
effect in nucleus-nucleus collisions. For $p-A$ collisions $d=1$ \cite{hwa02}.
If the physics for the longitudinal momentum degradation is the same in $p-A$ and
$A-A$ collisions, $d\simeq 1$ can be expected. A small deviation from $d=1$ in $A-A$
collisions may be due to the  multiple $NN$ collisions, each of which has a different
center of mass energy. Here, parameter $c$ is a measure of the strength of
longitudinal momentum degradation effect.

By changing variable from rapidity $y$ to pseudo-rapidity
$\eta$, using $p_L=p_T\sinh(\eta)$, one can get the particle distribution as follows
\begin{eqnarray}
\frac{dN}{2\pi p_Tdp_Td\eta}= \frac{Ap_T\cosh(\eta)}
{\sqrt{m_T^2+p_T^2\sinh^2(\eta)}}\nonumber\\
\exp(-\frac{\ln^2(\frac{m_T+p_T}{m}))}{2b}
-c(p_T\sinh(\eta))^{d})\ .
\label{maineq}
\end{eqnarray}
We assume that the four parameters $A$, $b$, $c$ and $d$ depend only on centrality
of the collisions at given collision energy, but not on the pseudo-rapidity.

\begin{figure}[tbph]
\includegraphics[width=0.45\textwidth]{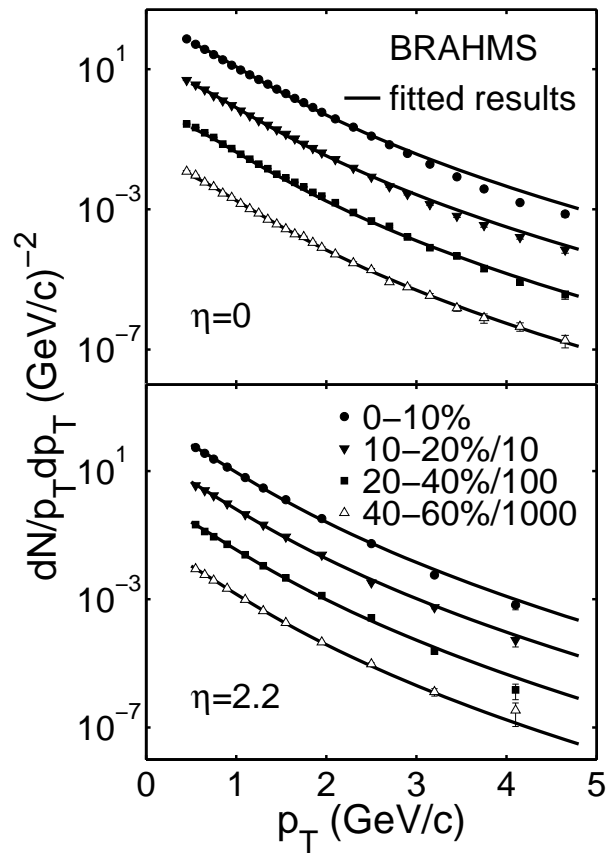}
\caption{The charged particle distribution $dN/2\pi p_Tdp_Td\eta$ as a function of
transverse momentum $p_T$ for Au+Au collisions at pseudo-rapidity
$\eta=0$ (top panel) and $\eta=2.2$ (bottom panel) for four centralities.
The data points are taken from \cite{data1}. The solid curves are the
fitted results with Eq. (\ref{maineq}). For clarity, the spectra
have multiplied by the indicated factors.}
\end{figure}

\section{Centrality dependence of particle distributions}
To determine the parameters for different centralities, we fit
the experimental data for produced particles with different centralities and
rapidities. For this purpose we choose the data from BRAHMS
Collaboration for Au+Au collisions at $\sqrt{s_{NN}}=200$ GeV
\cite{data1}. It needs to be mentioned that Eq. (\ref{maineq})
is supposed to be valid for a specific species of hadrons with given
mass $m$. In the chosen data, only the spectra of charged particles $(h^++h^-)/2$
are given, where all charged particles with different masses are included.
If the contribution to the spectra can be parameterized by Eq. (\ref{maineq})
for each species of charged hadrons, the distribution for charged particles
should be described by the sum of a few terms, each of which is given by Eq.
(\ref{maineq}). Then there would be too many parameters. Instead, in this paper,
we assume that Eq. (\ref{maineq}) is still valid for all charged particles, as long
as $m$ in Eq. (\ref{maineq}) is taken as an effective mass.
In principle, the effective mass should depend on the relative contents of all
charged particles. It is known that the proton to pion ratio is larger than 1 at
$p_T\sim 3 {\rm GeV}/c$ \cite{pr_pi} and that the kaon to pion ratio is about 0.2 in
a wide range of $p_T$. So the contribution to the distribution of charged particles
from protons and kaons will be quite large, especially for central collisions in the
mid-rapidity region. Thus the relative contents may be different for collisions at
different centralities. Even with the same centrality, the relative contents change
with $p_T$ and $\eta$. So generally speaking, the value of $m$ is a function of
centrality, $p_T$ and $\eta$. For simplicity, we assume in this paper that the value
of $m$ in Eq. (\ref{maineq}) is a constant for all centralities, $p_T$ and $\eta$.
For this reason, we choose, for different centralities and rapidities, the mass $m$
in Eq. (\ref{maineq}) to be 0.51 GeV, about the average mass of proton,
kaon and pion. With the charged particle $p_T$ distributions at $\eta=0$ and
2.2 at the same centrality cut, the fitting process can be completed in two steps.
We first fit the experimental distributions of transverse momentum at mid-rapidity
($\eta=0$) to determine the parameters $A$ and $b$, then, with fixed $A$ and $b$,
get parameters $c$ and $d$ from fitting transverse momentum distributions at $\eta=2.2$
at the same centrality. The experimental data chosen have a $p_T$ range from 0.45
to 4.65 GeV$/c$. We found from fitting that the value of $d$ is 0.94, very close
to 1 as expected, and almost the same for different centralities. So we fix $d=0.94$
for the distributions at other centrality cuts. The fitted results are shown in Fig.1
together with the experimental data from BRAHMS Collaboration for pseudo-rapidity
$\eta$=0 and 2.2 at four centralities. It is obvious that the data can be fitted very
well to the simple ansatz shown in Eq. (\ref{maineq}). The parameters obtained are
tabulated in TABLE I.

\tabcolsep 0.12in
\renewcommand\arraystretch{1.5}
\begin{table}
\begin{center}
\begin{tabular}{|c|c|c|c|c|}\hline
centrality & $A$ & $b$ & $c$ & $d$\\ \hline
${0}\sim{10}$\% & 239 & 0.351 & 0.0878 & 0.94\\ \hline
${10}\sim{20}$\% & 163 & 0.350 & 0.0826 & 0.94\\ \hline
${20}\sim{40}$\% & 93.3 & 0.344 & 0.0818 & 0.94\\ \hline
${40}\sim{60}$\% & 39.8 & 0.340 & 0.0788 & 0.94\\ \hline
\end{tabular}
\end{center}
\caption{Values of parameters $A$, $b$, $c$ and $d$ for distribution
of $dN/2\pi p_Tdp_Td\eta$ for four centrality cuts. The fitted data are from
\cite{data1}.}
\end{table}

In TABLE I the value of $b$ decreases slowly from central to peripheral collisions.
This means that $R_{CP}=(dN/N_Cp_Tdp_T)_C/(dN/N_Cp_Tdp_T)_P$ increases with $p_T$
at $\eta=0$. For pions, $R_{CP}$ decreases with $p_T$ for $p_T$ not too small, while
$R_{CP}$ for protons increases up to $p_T$ about 4 GeV$/c$ \cite{rcp}.
The behavior of $R_{CP}$ for unidentified hadrons is somewhat in between
those for pions and protons. The parameter $c$ is an
indicator of nuclear stopping in Au-Au collisions. The value of $c$ dictates the
difference between the spectra at different pseudo-rapidities at given centrality.
The larger the value of $c$, the bigger the difference, thus the stronger the nuclear
stopping effect. One sees that the nuclear stopping effect gets weaker for peripheral
collisions. This is in agreement with the naive expectation that the incident nucleons
need to traverse more nucleons in central collisions and lose more energy. Though the
values of $c$ are quite small, the little difference can give the difference of the $p_T$
distributions at different pseudo-rapidity $\eta$.

\begin{figure}[tbph]
\includegraphics[width=0.45\textwidth]{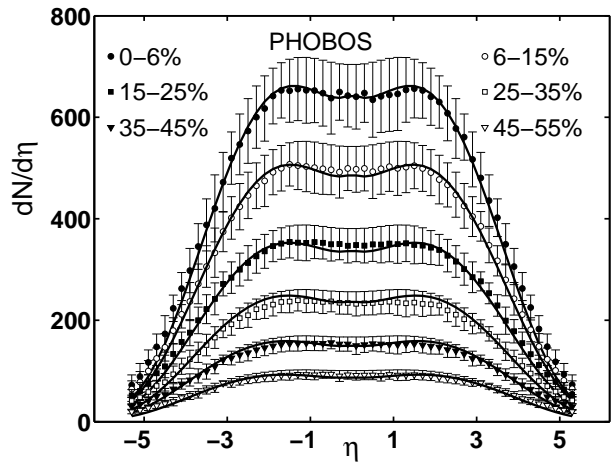}
\caption{Fitted invariant charged particle yields as a function of
pseudo-rapidity $\eta$ for six different centralities in Au+Au
collisions at $\sqrt{s_{NN}}=200$ GeV.  The data points are
taken from \cite{phobos}. Solid curves are from our fitting with Eq. (\ref{eq2}).}
\end{figure}

\tabcolsep 0.12in
\renewcommand\arraystretch{1.5}
\begin{table}
\begin{center}
\begin{tabular}
 {|c|c|c|c|c|}\hline centrality & $A$ & $b$ & $c$ & $d$\\ \hline
${0}\sim{6}$\% & 513.3 & 0.540 & 0.21 & 0.94\\ \hline
${6}\sim{15}$\% & 387.9 & 0.540 & 0.19 & 0.94\\ \hline
${15}\sim{25}$\% & 268.6 & 0.541 & 0.18 & 0.94\\ \hline
${25}\sim{35}$\% & 188.6 & 0.541 & 0.175 & 0.94\\ \hline
${35}\sim{45}$\% & 119.4 & 0.542 & 0.16 & 0.94\\ \hline
${45}\sim{55}$\% & 69.4 & 0.542 & 0.154 & 0.94\\ \hline
\end{tabular}
\end{center}
\caption{Values of parameters $A$, $b$, $c$ and $d$ for distribution
of $dN/d\eta$ for six centrality ranges. The fitted data are taken from \cite{phobos}.}
\end{table}

After obtaining the values of parameters $A, b, c$ and $d$ in
Eq. (\ref{maineq}) for the particle distribution, we can get the
pseudo-rapidity distribution for different centralities
\begin{equation}
\frac{dN}{d\eta}=2\int p_Tdp_T \frac{dN}{p_Tdp_Td\eta}\ ,
\label{eq2}
\end{equation}
where the factor 2 is included for the total number of all charged
particles. BRAHMS Collaboration have presented the pseudo-rapidity distributions
for 6 centralities \cite{data2}. Unfortunately, the centrality cuts in
\cite{data2} are different from those in \cite{data1}. So the obtained
parameters shown in TABLE I cannot be used for a comparison of our theoretical
expectation and the data on pseudo-rapidity distributions. Furthermore, we would
try to calculate the ratio between rapidity distributions for peripheral
and central collisions and compare it with experimental data which is given in
\cite{phobos} by PHOBOS Collaboration. So we now choose to fit the PHOBOS
data on rapidity distribution \cite{phobos}. One should keep
in mind that the pseudo-rapidity distribution of charged particles includes
contributions from charged particles with all possible transverse momentum
for any specific pseudo-rapidity. So soft particles with very low $p_T$
contribute most to $\eta$ distributions. Since the multiplicity of pions is always
much larger than those for protons and kaons when the transverse momentum is
integrated, pions always dominate
the pseudo-rapidity distribution. For this reason, the mass $m$ in Eq. (\ref{eq2})
should be very close to that for pions. So we take $m=m_\pi=0.139$ GeV in the calculation
of the pseudo-rapidity. Because the value of $m$ is different in fitting the $p_T$ and
$\eta$ distributions, other parameters will also be different. However, we will fix
$d=0.94$ in fitting the $\eta$ distributions. The data on pseudo-rapidity distribution
for the charged particle are over the pseudo-rapidity range $-5.4\leq\eta\leq 5.4$ for
six different centrality cuts. The fitted curves are shown in Fig. 2 together with the
data points. The agreement is excellent. The fitted parameters are tabulated in TABLE II.
The value of $b$ is almost constant for six centralities, but $c$ decreases faster from
central to peripheral collisions than shown in TABLE I. The small increase of parameter
$b$ from central to peripheral collisions is responsible for the decrease of the
suppression effect for pions in the collisions. Such a trend is opposite to that in
TABLE I, because in Fig. 1 protons contribute more in central collisions than in peripheral
ones and the production of proton is, instead of being suppressed, enhanced in central
collisions. When we focus on the pseudo-rapidity distributions, the effect of $c$ is
to narrow the distributions. In our model such effect is mainly from nuclear
stopping. The fact that the value of $c$ in TABLE II (where protons play less
important role) is much larger than that in TABLE I suggests that the nucleus stopping
effect can be illustrated more easily from pion spectra than from proton spectra.

\begin{figure}[tbph]
\includegraphics[width=0.45\textwidth]{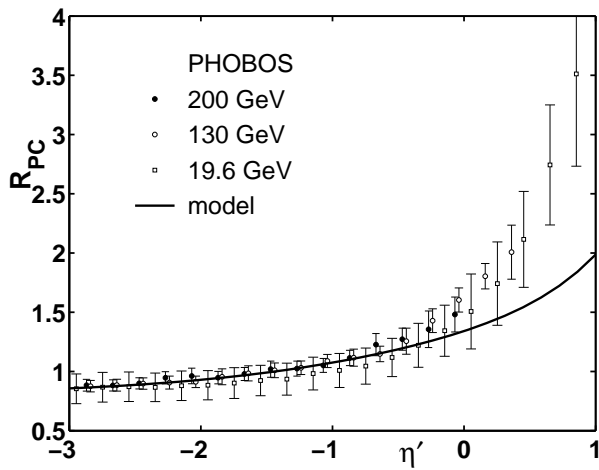}
\caption{The ratio $R_{PC}$ of $dN_{ch}/d\eta^\prime$ per participant pair between
peripheral (35-40\%) and central (0-6\%) collisions. The data are taken
from \cite{phobos}. The solid curve is for the calculated
results from parameterization Eq. (\ref{maineq}) and Eq. (\ref{eq2}).}
\end{figure}

One can go further to the very forward region with pseudo-rapidity $\eta\sim
y_{\rm beam}=5.36$ for $\sqrt{s_{NN}}=200$ GeV Au+Au
collisions to investigate the centrality dependence of rapidity distributions.
To compare the distributions at different colliding energies, a shifted variable is
introduced $\eta^\prime=\eta-y_{\rm beam}$, and an observable
$$R_{PC}=\left.\left(\frac{1}{N_p}\frac{dN}{d\eta}\right)_{35-40\%}\right/
\left(\frac{1}{N_p}\frac{dN}{d\eta}\right)_{0-6\%}$$
is measured in PHOBOS experiment at different colliding energies \cite{phobos}.
With the fitted parameters for the pseudo-rapidity distributions, the ratio $R_{PC}$
can easily be calculated. We use the parameters for centrality cut $35-45\%$ in
the calculation of $R_{PC}$, with $N_p$ given in \cite{phobos},
and the obtained results are plotted in Fig. 3 in
comparison with the data points with the centrality cut $35-40\%$ for the peripheral
collisions. Up to $\eta^\prime=0$ our calculated results agree with the data quite well.
For $\eta^\prime>0$, only data points at $\sqrt{s_{NN}}=19.6$ GeV are available. In
collisions at such a low energy, the contamination from the spectator nucleons to the
produced particles in the very forward region is considerably serious, especially
for peripheral collisions. This contamination will make $R_{PC}$ larger. That is
a possible origin of the discrepancy of our calculated results from the data at
large $\eta^\prime$. To verify this statement, more deliberate experimental
investigations are required.

\section{Conclusion}
In summary, we have investigated the transverse momentum and
pseudo-rapidity dependence of the distribution of charged particle
multiply for several centrality ranges, respectively. We assumed a factorized
parameterization for the invariant momentum distribution, with the transverse momentum
distribution at mid-rapidity being given by a Gaussian form of the
transverse rapidity and the longitudinal momentum distribution by a exponential
form characterizing the nuclear stopping effect. For the produced charged particles,
with suitably chosen effective mass, the simple phenomenological factorized form can
fit the experimental $p_T$ and $\eta$ distributions very well. Then the enhancement
of the rapidity distribution for peripheral collisions relative to central ones can be
explained naturally for very forward particle production. 

\acknowledgments{This work was supported in part by the National Natural Science
Foundation of China under grant No. 10475032 and by the Ministry of Education of
China under grant No. 03113.}


\begin{thebibliography}{99}
\bibitem{hwa84} R.C. Hwa, Phys. Rev. Lett. {\bf 52}, 492 (1984).

\bibitem{csernai} L.P. Csernai and J.I. Kapusta, Phys. Rev. {\bf D 29}, 2664 (1984).

\bibitem{huf} J. H\"ufner and A. Klar, Phys. Lett. {\bf B 145}, 167 (1984).

\bibitem{wong} C.Y. Wong, Phys. Rev. Lett. {\bf 52}, 1393 (1984).

\bibitem{kop} B.Z. Kopeliovich and B.G. Zakharov, Z. Phys. {\bf C 43}, 241
(1989); A. Capella and B.Z. Kopeliovich, Phys. Lett. {\bf B 381}, 325 (1996).

\bibitem{khar} D. Kharzeev, Phys. Lett. {\bf B 378}, 238 (1996).

\bibitem{van} S.E. Vance, M. Gyulassy, and X.-N. Wang, Phys. Lett. {\bf B 443},
45 (1998).

\bibitem{car73} M. Duong-van and P. Carruthers, Phys. Rev. Lett. {\bf 73},
133 (1973).

\bibitem{nao05} N. Suzuki and M. Biyajima, hep-ph/0510002,talk at XXXV International
Symposium on Multiparticle Dynamics 2005, Kromeriz, Czech Republuc, August 9-15, 2005;
hep-ph/0504076; hep-ph/0404112.

\bibitem{toma} B. Tomasik, Nucl. Phys. {\bf A 749}, 209 (2005) and references therein.
\bibitem{hwa02} R.C. Hwa and C.B. Yang, Phys. Rev. {\bf C 65}, 034905 (2002).

\bibitem{data1}I. Arsene et al., BRAHMS Collaboration, Phys. Rev. Lett.{\bf
91}, 072305 (2003)

\bibitem{pr_pi} S.S. Adler et al., PHENIX Collaboration, Phys. Rev.
{\bf C 69}, 034909 (2004)

\bibitem{rcp} F. Matathias, PHENIX Collaboration, J. Phys. G: Nucl. Part. Phys.
{\bf 30}, S1113 (2004).

\bibitem{data2}I.G. Bearden et al., BRAHMS Collaboration, Phys. Rev. Lett.{\bf
88}, 072305 (2002)

\bibitem{phobos} B.B. Back et al., PHOBOS Collaboration, Phys. Rev. Lett.
{\bf 91}, 052303 (2003).

\end{thebibliography}
\end{document}